\documentclass[onecolumn,showpacs,preprintnumbers,amsmath,amssymb,11pt,a4paper,showkeys]{revtex4}
\pdfoutput=1
\usepackage{geometry}
\usepackage{graphicx}
\usepackage{dcolumn}
\usepackage{bm}
\newcommand{\be}{\begin{equation}}
\newcommand{\ee}{\end{equation}}
\newcommand{\bd}{\begin{displaymath}}
\newcommand{\ed}{\end{displaymath}}
\newcommand{\BE}{\begin{eqnarray}}
\newcommand{\EE}{\end{eqnarray}}

\newcommand{\sgn}{{\rm sgn}}

\newcommand{\bx}{\ensuremath{\mathbf{x}}}

\newcommand{\bn}{\ensuremath{\mathbf{n}}}

\newcommand{\beff}{\ensuremath{\mathbf{f}}}

\newcommand{\avg}[1]{\left\langle{#1}\right\rangle}

\begin{document}

\preprint{}
\title{Imitation, internal absorption and the reversal of local drift in stochastic evolutionary games}


\author{Tobias Galla}
\email{tobias.galla@manchester.ac.uk}

\affiliation{Theoretical Physics, School of Physics and Astronomy, The University of Manchester, Manchester M13 9PL, United Kingdom}

\date{\today}

\begin{abstract}
Evolutionary game dynamics in finite populations is typically subject to noise, inducing effects which are not present in deterministic systems, including fixation and extinction. In the first part of this paper we investigate the phenomenon of drift reversal in finite populations, taking into account that drift is a local quantity in strategy space. Secondly, we study a simple imitation dynamics, and show that it can lead to fixation at internal mixed-strategy fixed points even in finite populations. Imitation in infinite populations is adequately described by conventional replicator dynamics, and these equations are known to have internal fixed points. Internal absorption in finite populations on the other hand is a novel dynamic phenomenon. Due to an outward drift in finite populations this type of dynamic arrest is not found in other commonly studied microscopic dynamics, not even in those with the same deterministic replicator limit as imitation.
\end{abstract}

\maketitle

\section{Introduction}
Much of the current interest in dynamic phenomena in the frequency-dependent selection in population focuses on the differences between stochastic descriptions of evolutionary processes and their deterministic counterparts. Traditionally deterministic formulations have been more common in the biological literature, presumably due to the fact that the theory of non-linear deterministic differential equations is relatively well developed. Aspects of stochastic dynamics that have been studied in recent years include (i) probabilities with which mutants invade existing populations \cite{nowakbook}, (ii) fixation probabilities and time-to-fixation in finite populations \cite{nowakbook,altrockinger}, (iii) the differences between dynamical stationary states of deterministic and stochastic processes \cite{traulsen, traulsenreview, mobilia} and (iv) the dynamics of pattern formation induced by noise \cite{frey}.

In the present paper we will focus on two dynamic features of stochastic evolutionary dynamics, namely those of (a) drift reversal and (b) fixation to internal points in strategy space. Drift reversal here refers to a phenomenon reported in \cite{traulsen2, claussen2,claussen1}, and characterises the observation that some stochastic evolutionary systems tend to approach the boundaries of the strategy space at small population sizes, whereas a drift in the other direction, towards the centre of strategy space, occurs at larger sizes. In particular expressions for a critical population size, at which this reversal of drift occurs, were derived in \cite{traulsen2, claussen2,claussen1}. It is important to stress that the meaning of the word `drift' is here different from its usual use for example in the context of `random drift', `genetic drift' or `neutral drift' \cite{kimura,kimura2,hubbell}. Instead it measures a tendency of the stochastic system to approach an internal fixed point, or to approach the boundaries of strategy space respectively, this will become clear below, when we give a more precise mathematical characterisation.

Secondly we analyse the dynamics of a strict proportional imitation process in finite populations, the details of this dynamics will be described below. Game dynamical processes commonly described in the literature are typically of the following type: in a first step two individuals are drawn at random from the population. Secondly, if the two individuals are of distinct types, one individual adopts the strategy of the other with a certain rate, typically dependent on the fitnesses of the two individuals, and in some cases also on the mean fitness of all individuals in the population. This adaptation step is equivalent to the death of one individual, and the birth of another. This type of dynamics leads to two distinct sources of stochasticity, one coming from the selection of the two individuals, and a second from the randomness of the birth and death processes that occur subsequently. In \cite{traulsendet} an interesting limiting case was introduced and studied. Both types of randomness were eliminated, leading to an entirely deterministic process even in finite populations. Here we take a different route, and make the birth-death component of the update deterministic in that only transition towards strictly more successful strategies are allowed. The randomness in the selection of individuals for potential update remains however, and consequently the overall dynamics is still stochastic, but as we will see novel phenomena can occur, in particular the possibility of an absorbing fixed point in the interior of the strategy space.

The paper is organised as follows: In Sec. \ref{sec:def} we will introduce the required definitions and the general notation. In Sec. \ref{sec:dr1d} we then analyse the phenomenon of drift reversal in detail, and provide a refinement of the findings of \cite{traulsen2, claussen2,claussen1}. These are derived for the one-dimensional case of symmetric games with two strategies, but can be generalised to higher-dimensional dynamics. In Sec. \ref{sec:abs1d} we then introduce the strict imitation dynamics, and show how it leads to internal absorption in the context of stochastic one-dimensional dynamics. Sections  \ref{sec:imitrps} and \ref{sec:imitbs} and then focus on similar dynamics for cyclic games with three strategies per player and for two-population games. In the final section we summarise our findings.


\section{General definitions}\label{sec:def}
\subsection{Stochastic dynamics in finite populations}
We will consider populations of $N$ agents, and restrict the analysis to processes conserving the size of the population. Each individual can be of one of $S$ types (the different species or pure strategies), so that the state of the system is characterised by the vector $\bn=(n_1,\dots,n_{S})$, where $n_i$ indicates the number of individuals of type $i\in\{1,\dots,S\}$. One always has $\sum_{i=1}^{S}n_i=N$.  For one-population models we will label the payoff matrix elements of the underlying game as $a_{ij}$, where $i,j\in\{1,\dots,S\}$. The quantity $a_{ij}$ thus denotes the payoff a player of type $i$ receives in interaction with a player of type $j$. Frequency-dependent fitnesses are then given by
\be
\pi_i(\bn)=\frac{1}{N-1}\left(\sum_{j}a_{ij}n_j-a_{ii}\right),
\ee
where we have excluded interaction of one individual with itself.
To specify a population dynamical process it is then sufficient to define 
the `conversion' rates $T_{i\to j}$, corresponding to events in which 
a player of type $i$ is replaced by one of type $j$. We will limit the discussion to 
processes of the general form
\be
T_{i\rightarrow j} = \frac{n_i}{N}\frac{n_j}{N}g_{ij}(\beff), ~~ i,j\in\{1,\dots,S\}, ~~ i\neq j,
\label{eq:transition}
\ee
where $f_i=1-w+w\pi_i$ and $\beff = (f_1, \dots, f_S)$. The model parameter $w$ here denotes the strength of selection and takes values $w\in[0,1]$. The $\{f_i\}$ are often referred to as reproductive fitnesses. In this paper we will mostly consider pairwise comparison processes, and the precise choice of $w$ will often be unimportant, amounting essentially to choosing a time scale. The form (\ref{eq:transition}) is found
by, at each time step, selecting two players from the population (with replacement), one for potential reproduction 
and one for potential removal. The player selected for potential removal is assumed to be of type $i$, the one for potential reproduction of type $j$. For a given pair of selected  players reproduction and death actually only occur at a rate proportional 
to $g_{ij}(\beff)$.
The time evolution of the probability density $P(\bn,t)$ of finding the system in state $\bn$ at time $t$ is then given by
\be
\label{eq:master}
\frac{dP(\bn,t)}{dt} = \sum_{i\neq j} (\widehat E_i^{}\widehat E_j^{-1}-1) \left[T_{i\to j}(\bn)P(\bn,t)\right],
\ee
where $\widehat E_i$ is a shift operator, acting on functions of $\bn$ by increasing $n_i$ by one: $\widehat E_i \psi(n_1,\dots,n_S)=\psi(n_1,\dots,n_{i-1},n_i+1,n_{i+1},\dots,n_S)$. $\widehat E_i^{-1}$ is the inverse operator, reducing $n_i$ by one.
It is here important to note that the system only has $S-1$ degrees of freedom, given that $\sum_{i=1}^Sn_i=N$. For reasons of notational compactness the above master equation (\ref{eq:master}) has been formulated for the full system $\bn=(n_1,\dots,n_S)$. This equation will be the starting point for our further investigations of stochastic game dynamics in finite populations.
\subsection{Deterministic replicator dynamics}
A description of evolutionary processes for infinite populations, $N\to\infty$, is given by the following replicator dynamics
\be\label{eq:repl0}
\dot x_i=x_i(f_i^\infty(\bx)-\phi^\infty(\bx)), ~~ i=1,\dots,S.
\ee
Here $x_i=\lim_{N\to\infty} \frac{n_i}{N}$ is the fraction of players of type $i$ in the limiting infinite population, and $\phi^\infty$ the average fitness in the population
\be
\phi^\infty=\sum_i x_i f_i^\infty(\bx).
\ee
It is however important to note that the above replicator equation provides an accurate description of the dynamics in the limit $N\to\infty$ only for a suitable choice of the microscopic dynamics, in particular of the functions $g_{ij}$ defined above. This will be explained in more detail below, see also \cite{traulsen2, bladon} for further discussion. Throughout this paper the superscript $\infty$ will indicate that we are referring to quantities defined for infinite populations. These quantities will typically be functions of $\bx=(x_1,\dots,x_S)$, and can be obtained by a simple replacement of $\lim_{N\to\infty} n_i/N$ by $x_i$, in the corresponding quantities for finite populations. For example we have $f_i^\infty(\bx)=1-w+w\sum_j a_{ij} x_j$.


\section{Reversal of local drift in $2\times 2$ symmetric games}\label{sec:dr1d}
\subsection{Definitions}
We will first consider the example of a two-player symmetric game with two strategies, which we will label $i\in\{A,B\}$. In particular we will address games for which the deterministic replicator dynamics has a stable internal fixed point. Specifically we will choose the so-called Hawk-Dove game, also referred to as the game of chicken, defined by the payoff matrix \cite{nowakbook}
\be
\left(\begin{array}{cc}(b-c)/2 & b \\
0 & b/2\end{array}\right),
\ee
with model parameters $b,c>0$, to be specified below. Here $S=2$ and the state of the population is fully specified by the number of individuals carrying strategy $A$. If the population consists of $n$ players of type $A$, and $N-n$ players of type $B$ one then has
\BE
\pi_A(n)&=&\frac{n-1}{N-1}\frac{b-c}{2}+\frac{N-n}{N-1}b,\nonumber\\
\pi_B(n)&=&\frac{N-n-1}{N-1}\frac{b}{2}.
\EE
The number $n^*$ of individuals of type $A$ at which the fitnesses of the two species are equal, $\pi_A(n^*)=\pi_B(n^*)$ is given
\be\label{eq:nstar}
n^*=\frac{b}{c}N+1,
\ee
if this expression is integer-valued (otherwise there is not such point). We will in the following specifically choose $b=1$ and $c=2$, so that $n^*=N/2+1$. This expression is integer-valued whenever the total number of individuals in the population, $N$, is even, and we will restrict the discussion to such cases in the following.

In order to discuss the phenomenon of local drift reversal we will focus on what is referred to as the so-called `local' update process in the literature \cite{traulsen2,claussen2,claussen1}. This process is defined by the following transition rates
\BE
T^+(n)\equiv T_{B\to A}(n)&=&\frac{n}{N}\frac{N-n}{N}\left(\frac{1}{2}+\frac{w}{2}\frac{\pi_A(n)-\pi_B(n)}{\Delta\pi_{\mbox{\tiny max}}}\right),\\
T^-(n)\equiv T_{A\to B}(n)&=&\frac{n}{N}\frac{N-n}{N}\left(\frac{1}{2}-\frac{w}{2}\frac{\pi_A(n)-\pi_B(n)}{\Delta\pi_{\mbox{\tiny max}}}\right).
\EE
In this two-strategy case we write $T^+$ as a short-hand for $T_{B\to A}$ and $T^-$ for the transition rate in the opposite direction. $\Delta\pi_{\mbox{\tiny max}}$ is an additional model parameter, introduced to ensure transition rates remain positive. Specifically this model parameter denotes the maximally possible difference in fitness between the fittest and the least fit species \cite{traulsen2}. Given the choices $b=1,c=2$ we will here use $\Delta\pi=w=1$.

The deterministic dynamics corresponding to the local update rule, valid in the limit of infinite populations, $N\to\infty$ can be obtained from these definitions for example from the lowest order terms of an expansion of the master equation (\ref{eq:master}) in the square root of the inverse system size. We will not present the mathematical details here, as these have been discussed at length in the literature \cite{bladon, traulsen2,kampen}. The lowest-order result is obtained as
\be\label{eq:repl2}
\dot x=T^{\infty,+}(x)-T^{\infty,-}(x)=\frac{w}{2\Delta\pi_{\mbox{\tiny max}}}x(1-x)(1-2x),
\ee
where $x$ is the fraction of players of type $A$ in the limit of infinite populations. It is obvious that this equation admits an internal fixed point at $x^*=1/2$ in addition to the trivial fixed points $x^*=0$ and $x^*=1$. Up to a re-scaling of time equation (\ref{eq:repl2}) is a specific example of a replicator equation (\ref{eq:repl0}), as some elementary algebra shows.
\subsection{Inversion of local drift}
\begin{figure}
\vspace{2em}
\centering
\includegraphics[scale=0.6]{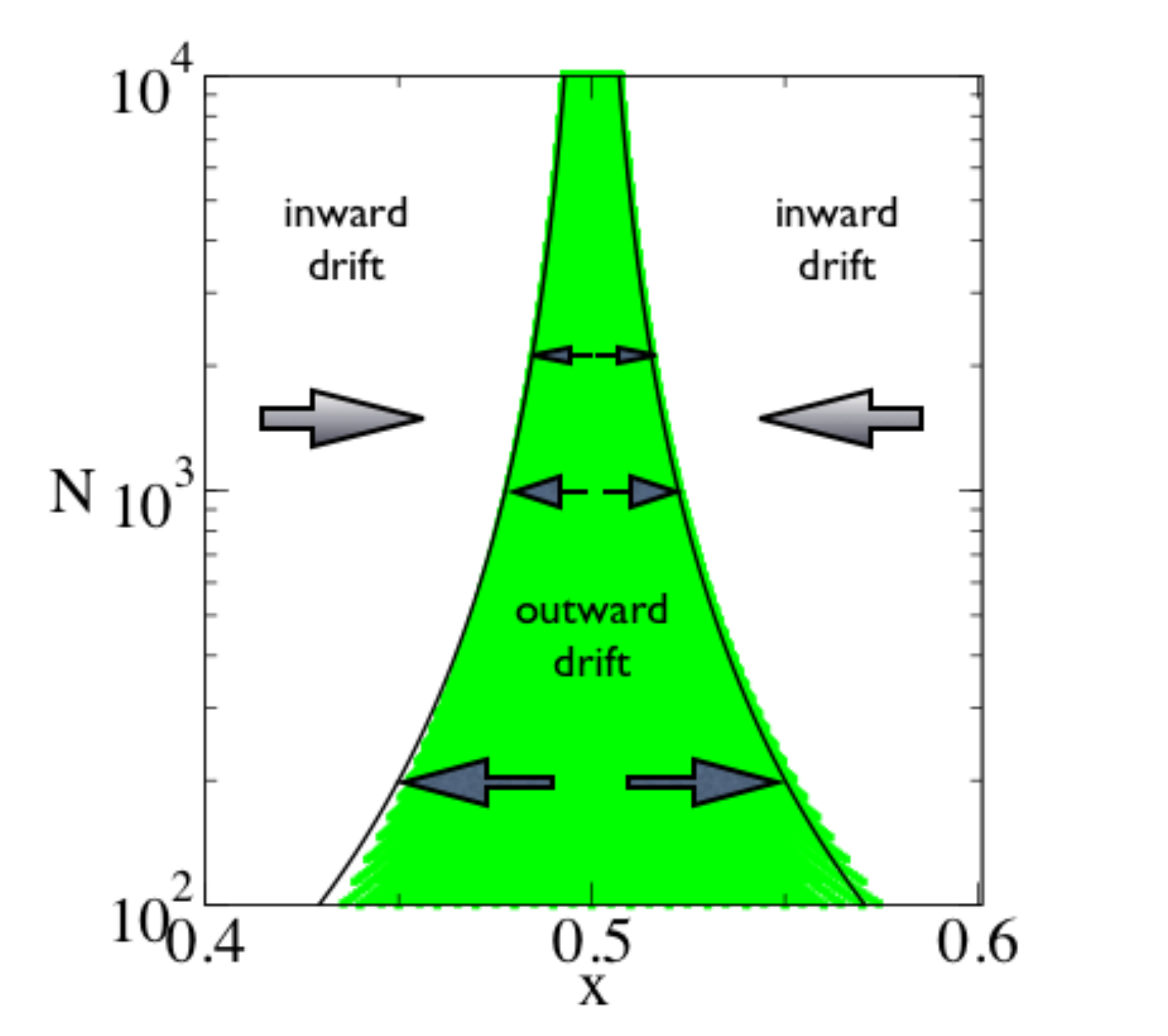}
\caption{(Color online) Dependence of the direction of local drift in the Hawk-Dove game as a function of the starting point $x=n/N$ and of the system size $N$ of the finite population. An outward drift occurs in a region close to the deterministic fixed point $x^*=1/2$ (shaded area, results are obtained from Eq. (\ref{eq:drift0})), and inward drift further away from the deterministic fixed point. The range of starting points at which an outward drift is observed is reduced as the system size is increased. In infinite populations the drift is inwards throughout strategy space, with $x^*=1/2$ a globally stable attractor. Black lines show the estimate of Eq. (\ref{eq:driftest}).}
\label{fig:drift1d2}
\end{figure}

The notion of drift in \cite{traulsen2, claussen2,claussen1} refers to the tendency of the stochastic dynamics to evolve towards an internal fixed point, or away from it. More formally, given a measure of distance of the one-dimensional Hawk-Dove system from the fixed point, $D^\infty:[0,1]\longrightarrow {\mathbb R}_0^+$, we will ask whether this distance is more likely to increase than to decrease if the system is started at a certain point in phase space, or whether there is a higher probability for a reduction in distance than there is for an increase of distance. $D^\infty$ may here for example be the Euclidean distance $D^\infty(x)=(x-1/2)^2$ from the internal fixed point $x^*=1/2$ of the deterministic dynamics, but for purposes of general considerations we will leave $D^\infty(\cdot)$ unspecified until a later point.  Recall that following our above remarks we will use the notation $D^\infty(x)$ to refer to the case of infinite populations, and $D(n)\equiv D^\infty(n/N)$ for $N$ finite.
\begin{figure}[t]
\vspace{2em}
\centering
\includegraphics[scale=0.6]{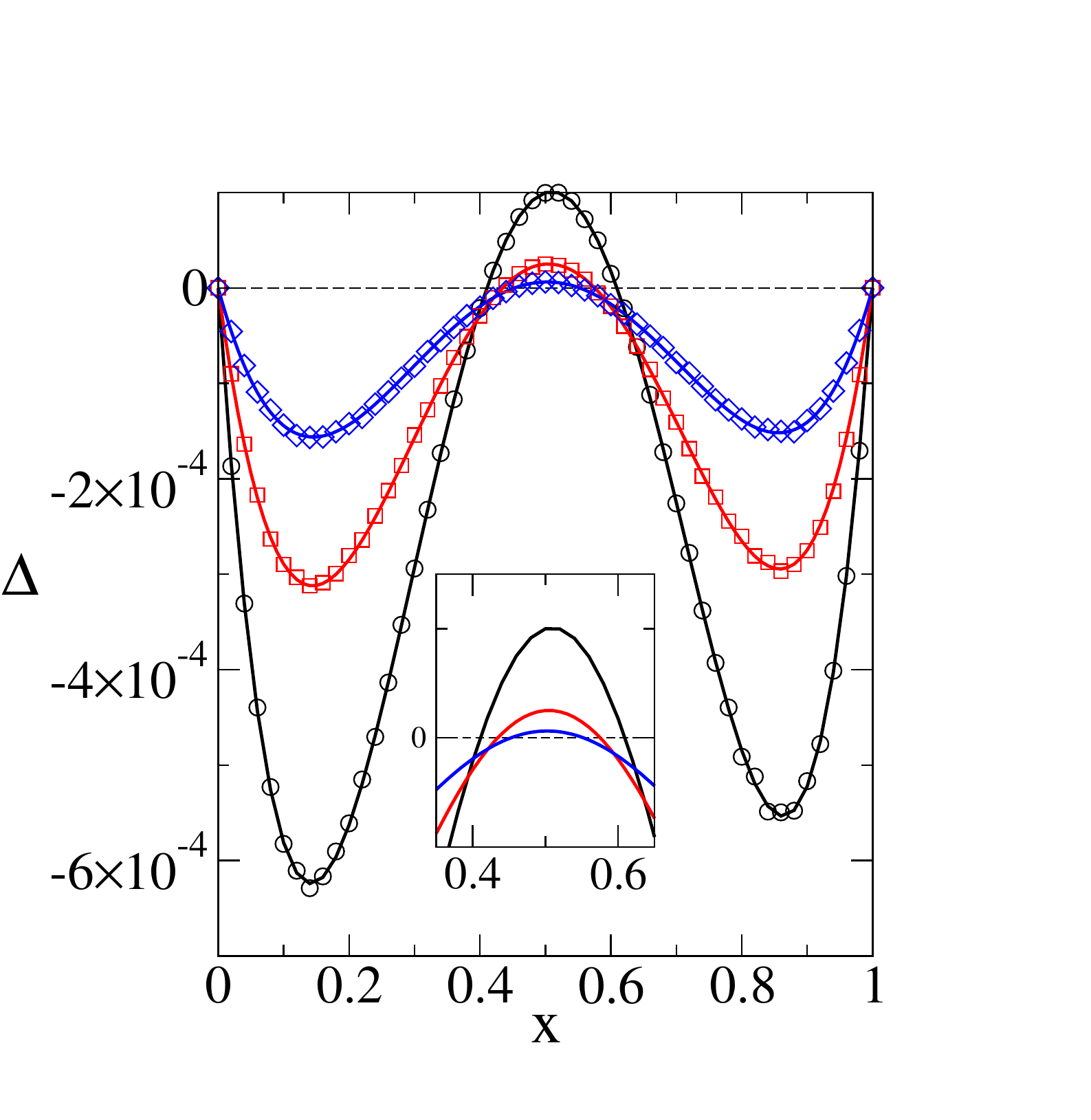}
\caption{(Color online) Local drift in the Hawk-Dove game. Lines are from theory ($N=50,100,200$ from top to bottom at the maximum), symbols from simulations (averaged over $10^6$ independent runs). Inset is a magnification of the central peak, where an outward drift is observed.}
\label{fig:drift1d}
\end{figure}
The expected change $\Delta(n)$ of the distance of the system from the fixed point, conditioned on a starting point $x=n/N$, is then given by
\BE\label{eq:delta}
\Delta(n)&\equiv&T^+(n) D(n+1)+T^-D(n-1)-(T^+(n)+T^-(n)D(n)\nonumber\\
&=& \frac{T^+(n)-T^-(n)}{N}(D^\infty)'(x)+\frac{T^+(n)+T^-(n)}{2N^2}(D^\infty)''(x)+{\cal O}(N^{-3}),
\EE
where $(D^\infty)'(x)$ and $(D^\infty)''(x)$ refer to the first and second derivatives with respect to $x$ of the function $D^\infty(x)$. Focus sing now on the Euclidean distance $D^\infty(x)=(x-1/2)^2$ we then have
\BE\label{eq:drift0}
\Delta(n)&=&2\frac{T^+(n)-T^-(n)}{N}\left(\frac{n}{N}-\frac{1}{2}\right)+\frac{T^+(n)+T^-(n)}{N^2}.
\EE
For the local process as specified above we find
\be\label{eq:drift1}
\Delta(n)=\frac{n(N-n)}{N^2}\frac{1}{(N-1)}\left[1-\frac{2(n-1)}{N}\right]\left(\frac{n}{N}-\frac{1}{2}\right)+\frac{n(N-n)}{N^2}\frac{1}{N^2}.
\ee
This result is illustrated and compared with numerical simulations in Fig. \ref{fig:drift1d}. Simulations are of the microscopic process are here carried out using the popular Gillespie algorithm \cite{gillespie}. As seen in the figure the theoretical result (\ref{eq:drift0}) is confirmed by these computer experiments. Results show that for any fixed system size $N$, a local outward drift, $\Delta>0$ occurs for starting points, $x=n/N$, close to the deterministic fixed point. If the starting point is sufficiently far away from the fixed point an inward drift occurs instead. As demonstrated in the inset of Fig. \ref{fig:drift1d} the region in which an outward drift occurs shrinks consistently as the system size is increased. This is illustrated further in Fig. \ref{fig:drift1d2}, where we show where in strategy space inward and outward drifts occur as a function of system size. In particular we notice that drift is a local quantity, at a given system size and at fixed model parameters the drift may be inward for one starting point in strategy space, but outward in other regions. The range of starting points at which an outward drift is observed is reduced as the system size is increased. In infinite populations the drift is inward throughout strategy space, with $x^*=1/2$ a globally stable attractor.

An analytical estimate for the boundary separating the regions with inward and outward drifts can be obtained setting $\Delta(n)=0$ in Eq. (\ref{eq:drift1}). Re-arranging and ignoring sub-leading terms in higher powers of $N^{-1/2}$ one finds that a reversal of local drift occurs when
\be\label{eq:driftest}
\frac{n}{N}=\frac{1}{2}\pm\frac{1}{\sqrt{2N}},
\ee
i.e. when $(n/N-1/2)^2=1/(2N)$, establishing a typical distance $\overline{D}=1/(2N)$ from the deterministic fixed point. Eq. (\ref{eq:driftest}) thus indicates the point in strategy space at which, for a given system size, the reversal of local drift occurs. Conversely for a given point $x=n/N$ in strategy space the reversal of local drift is found at a population size $N$ determined by $N=[\sqrt{2}(x-1/2)]^{-2}$.

These results can be connected to the well established system-size expansion approach to evolutionary processes in finite populations. We here discuss an approach based on van Kampen's system size expansion, an alternative line of reasoning, using a Kramers-Moyal expansion is outlined in the Appendix. Following \cite{kampen, bladon} the following stochastic differential equation for fluctuations about the deterministic fixed points can be derived
\be\label{eq:vk1}
\dot \xi(t)=-\frac{1}{4}\xi(t)+\eta(t).
\ee
The variable $\xi$ is here defined as the re-scaled deviation 
\be\label{eq:vk2}
\xi=\sqrt{N}\left(\frac{n}{N}-\frac{1}{2}\right)
\ee
from the deterministic fixed point, the noise variable $\eta(t)$ is Gaussian and uncorrelated in time, specifically one obtains 
\be\label{eq:vk3}
\avg{\eta(t)\eta(t')}=\frac{1}{4N}\delta(t-t')
\ee
from the van Kampen expansion. We will not report the details of the derivation of Eqs. (\ref{eq:vk1},\ref{eq:vk2},\ref{eq:vk3}), these follow the steps of \cite{kampen,bladon}. It is however worth noting that these equations apply only in the long-time limit, when the deterministic dynamics have reached a fixed point. The van Kampen expansion can also be executed for transients or deterministic systems with attractors other than fixed points \cite{boland}, but this is not relevant for the purposes of the present paper.

Eqs. (\ref{eq:vk2},\ref{eq:vk3}) define a linear Langevin equation, and it is hence straightforward to compute the variance of fluctuations in the stationary state, see e.g. \cite{risken} for details. One finds that $\frac{d}{dt}\avg{\xi(t)^2}=-\frac{1}{2}\avg{\xi(t)^2}+\frac{1}{4N}$, and hence $\avg{\xi(t)^2}=\frac{1}{2N}$ at long times $t$. This defines a `typical' asymptotic distance from the deterministic fixed point in perfect agreement with Eq. (\ref{eq:driftest}). We therefore conclude that the stochasticity of the dynamics induces a typical asymptotic distance $\overline D=1/(2N)$ from the deterministic fixed point, and that an inward drift is observed for starting points in strategy space that are further away from the fixed point than this typical distance. An outward drift is found for starting points closer to the fixed point than the `preferred' asymptotic distance, underlining the importance to think of drift as a local quantity. In \cite{claussen1,claussen2} similar calculations were performed (in the context of different games), but a flat average over all possible initial conditions was then computed, resulting in a global notion of an inward or outward drift, depending solely on the system size $N$, but with no `local' character in strategy space. While expressions for the local drift were of course contained in the calculations of  \cite{claussen1,claussen2} as intermediate steps, we believe that our results and analysis provide an interesting addition and further insight into the dynamics of drift reversal.


\section{Imitation dynamics and internal absorption in one-dimensional systems}\label{sec:abs1d}
We now turn to the second main topic of this paper, and consider a microscopic update process which admits absorption at an internal fixed point even for finite populations. This is not the case in the previous section, where we have shown that finite systems under the local update rule instead tend to a stationary state with fluctuations of a typical magnitude, or distance from the fixed point, even if the corresponding deterministic dynamics have a stable fixed point within the interior of strategy space.

Specifically we will now consider an imitation process. As before two players are drawn at random from the population at the beginning of each microscopic update step. If they are of identical types, nothing happens. If one player is of type $A$ and the other one of type $B$, then the two players compare their fitnesses, $f_A$ and $f_B$. If the two fitnesses are equal, $f_A=f_B$, the player of type $A$ converts into a player of type $B$ with a rate proportional to $\nu/2$, and similarly $B$ switches to $A$ with rate $\nu/2$. The non-negative constant $\nu$ is a model parameter, and denotes the rate with which `neutral' transitions occur.  If however one player has a strictly higher fitness than the other, then the player with the higher fitness remains unchanged. The player with the lower fitness adopts the higher-performing strategy with a rate proportional to the fitness difference, for example if $f_A>f_B$, then player $B$ will turn into a player of type $A$ with a rate proportional to $f_A-f_B$.  No transitions to strategies with strictly lower fitness are allowed. Specifically we will consider the process defined by the following transition rates:
\BE
T^+(n)&=&\frac{n}{N}\frac{N-n}{N}\left(\frac{\nu}{2}+\frac{w}{2}\frac{\pi_A(n)-\pi_B(n)}{\Delta\pi_{\mbox{\tiny max}}}\right)\Theta(\pi_A(n)-\pi_B(n))\nonumber \\
T^-(n)&=&\frac{n}{N}\frac{N-n}{N}\left(\frac{\nu}{2}-\frac{w}{2}\frac{\pi_A(n)-\pi_B(n)}{\Delta\pi_{\mbox{\tiny max}}}\right)\Theta(\pi_B(n)-\pi_A(n))
\EE
$\Theta(\cdot)$ is the step function, i.e. we have $\Theta(y<0)=0$, $\Theta(y\geq 0)=1$. \begin{figure}
\vspace{2em}
\centering
\includegraphics[scale=0.4]{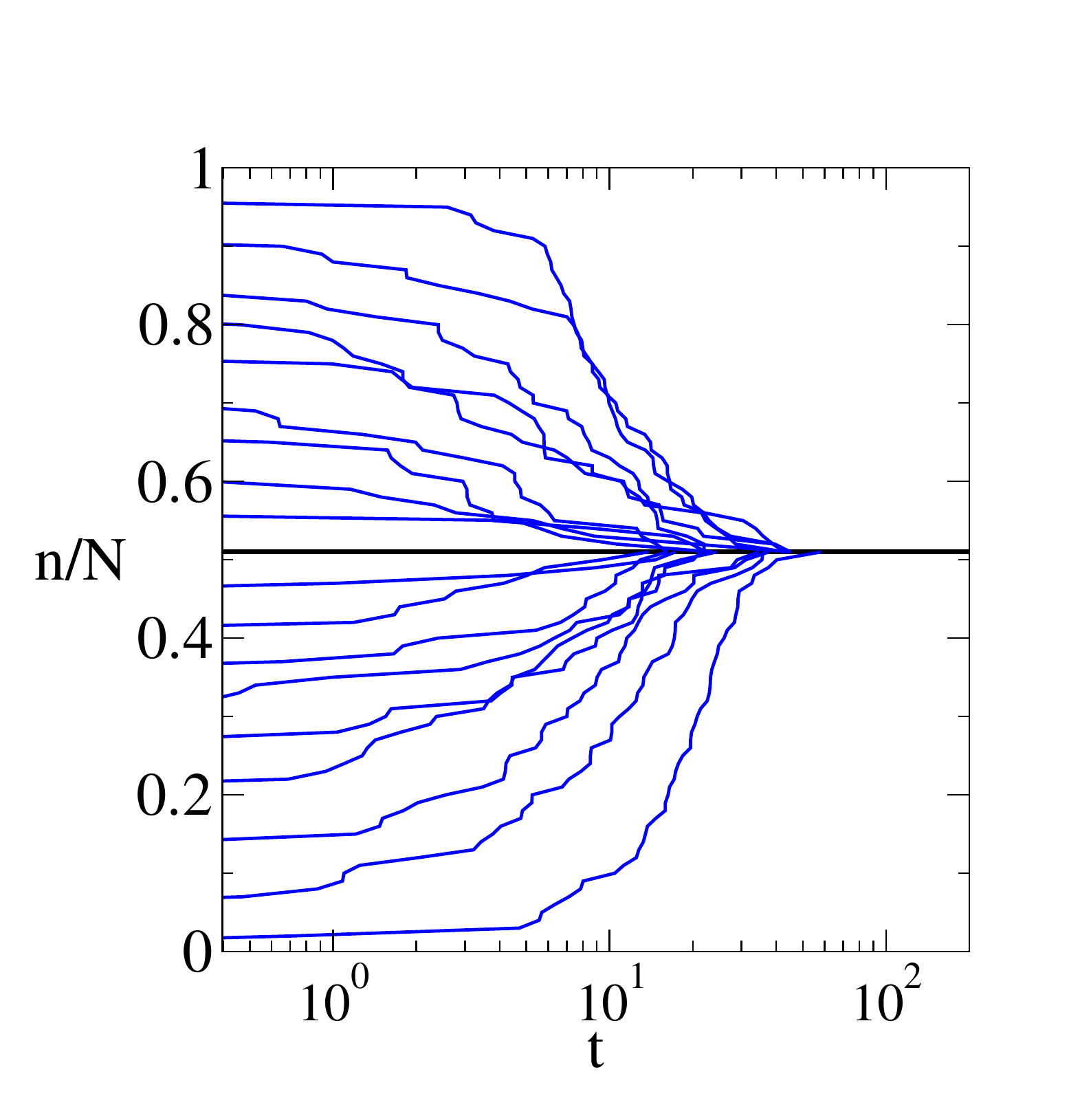}~~~\includegraphics[scale=0.4]{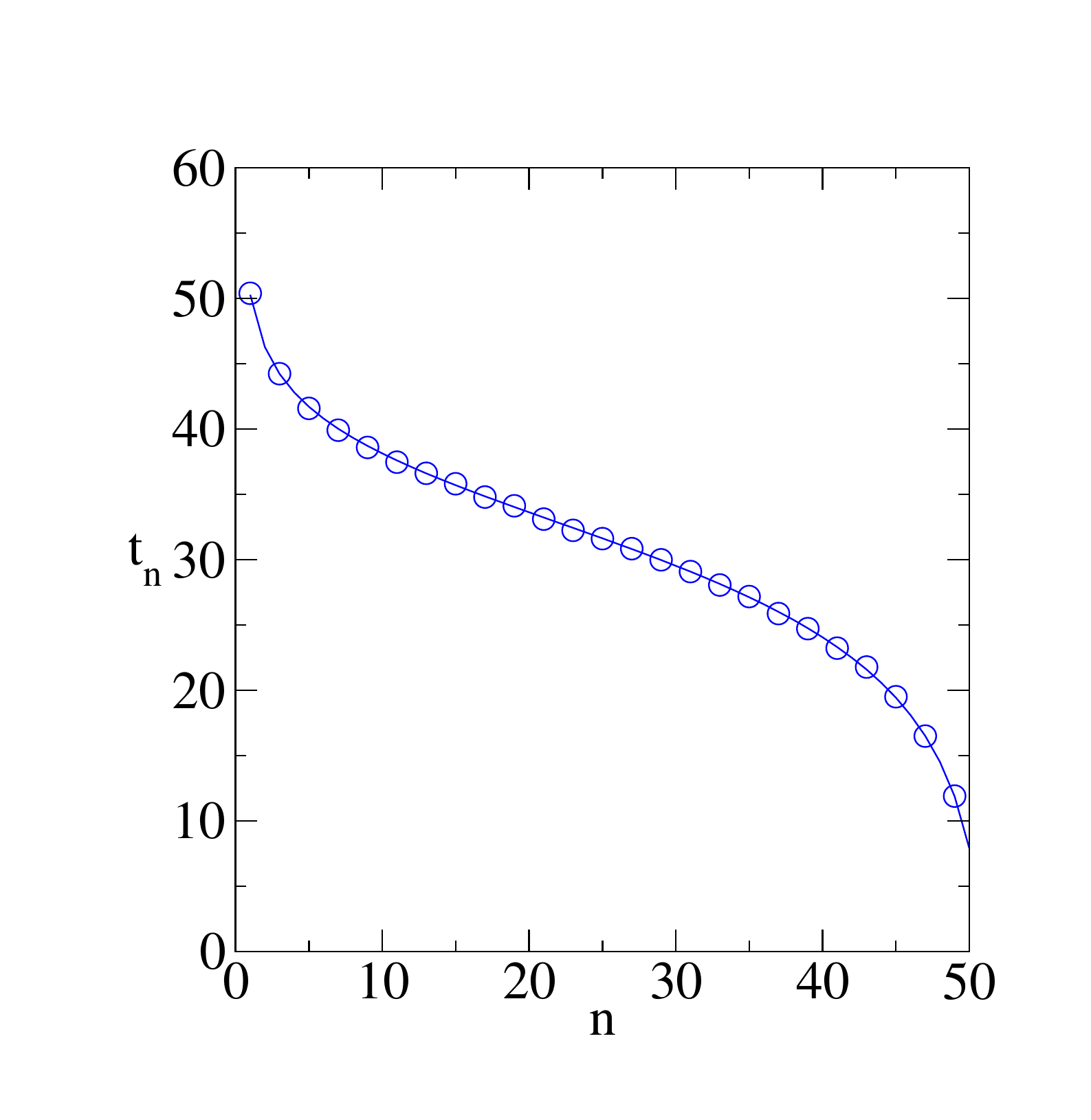}
\caption{(Color online) Internal absorption in the imitation dynamics of the Hawk-Dove game. Left: Trajectories for $N=100$ players, started from different initial conditions. The horizontal line shows the internal absorbing state, $n^*/N=51/100$, as defined by Eq. (\ref{eq:nstar}). Right: Average time to fixation at the internal fixed point in a population of $N=100$ agents, started at point $n<n^*=51$ in strategy space (solid line are results from Eq. (\ref{eq:ttf}), interpolated to non-integer $n$ for optical convenience, symbols are from simulations, averaged over $10000$ independent runs).}
\label{fig:absorb1d}
\end{figure}
The deterministic limit of this dynamics is obtained as
\BE
\dot x&=&T^{\infty,+}(x)-T^{\infty,-}(x) \nonumber \\
&=&x(1-x)\left(\frac{\nu}{2}\sgn(\pi_A(i)-\pi_B(i))+\frac{w}{2}\frac{\pi_A(i)-\pi_B(i)}{\Delta\pi_{\mbox{\tiny max}}}\right),
\EE
where $\sgn(y<0)=-1, \sgn(y=0)=0$ and $\sgn(y>0)=1$. Inserting the expressions for $\pi_A$ and $\pi_B$ of the Hawk-Dove game ($b=1, c=2$) we have
\be\label{eq:detimi}
\dot x =x(1-x)\left(\frac{\nu}{2}\sgn(1-2x)+\frac{w}{4}\frac{1-2x}{\Delta\pi_{\mbox{\tiny max}}}\right).
\ee
We will here mostly focus on the case $\nu=0$, which we will refer to this as strict imitation, or strict proportional imitation. In this case the choice of $w$ and $\Delta\pi_{\mbox{\tiny max}}$ is mostly irrelevant, and equivalent to setting the time scale. Unless specified otherwise we will mostly use $\Delta\pi_{\mbox{\tiny max}}=w=1$ in the following. The word `proportional'  indicates that adaptation towards better strategies occurs with a rate proportional to the improvement in fitness.  As first shown in \cite{helbing} one recovers the standard replicator dynamics for $\nu=0$, see also \cite{hofbauer,gintis}. Indeed Eq. (\ref{eq:detimi}) reduces to  Eq. (\ref{eq:repl2}) for $\nu=0$ up to an overall re-scaling of time by a constant factor $1/2$.

Even though the local process discussed in the previous section and the strict imitation dynamics have identical deterministic limits the dynamical features of the respective stochastic processes in finite populations are quite different. This is illustrated in Fig. \ref{fig:absorb1d}, where we show a set of trajectories obtained from the strict imitation dynamics ($\nu=0$) for the Hawk-Dove game ($b=1,c=2$), started at different initial conditions. As seen in the figure all trajectories are eventually absorbed at the internal fixed point given by Eq. (\ref{eq:nstar}). 

This behaviour can be understood from considerations of local drift taking into account that at $\nu=0$ we have $T^+(n)T^-(n)=0$ for all $n\in\{0,\dots,N\}$, i.e only one of the two transition rates is ever non-zero for any given $n$. If for example $n>N/2+1$, then we have $f_A(n)<f_B(n)$, and hence $T^+(n)=0$. Excluding the trivial case $n=N$, which is an absorbing state itself, the resulting drift is then given by
\be
\Delta(n)=T^-(n) \left[D(n-1)-D(n)\right]<0,
\ee
with the last inequality holding as $T^-(n)>0$ and $D(n)>D(n-1)$ for $n>N/2+1$. A similar argument applies when $0<n<N/2+1$. At the fixed point $n^*=N/2+1$ we have $f_A(n^*)=f_B(n^*)$, and therefore $T^+(n^*)=T^-(n^*)=0$, confirming that $n^*=N/2+1$ is indeed an absorbing state. One therefore concludes that fixation at the internal attractor will occur for any non-trivial starting point (i.e. $n\neq0, ~ n\neq N$), and provided that $N$ is an even number (ensuring that $n^*=N/2+1$ is an integer).

To complete the picture it remains to compute the mean time to fixation, given a starting point $n$. Assume to this end that $n<n^*$. Then $T^-(n)=0$, and the system will move to state $n+1$ with the rate $T^+(n)$, reaching the state $n+1$ after an average waiting time of $1/(T^+(n))$. If $n+1<n^*$ the system has not yet reached absorption. A second exponential clock will start ticking, and the system will hop to state $n+2$ after an average waiting period $1/(T^+(n+1))$. This procedure then iterates until the absorbing state $n^*$ is reached, and we conclude that the mean time to fixation for trajectories starting at $n<n^*$ is given by
\be\label{eq:ttf}
\tau_n=\sum_{k=n}^{n^*-1}\frac{1}{T^+(k)}.
\ee
Similar results can be derived for starting points $n>n^*$. These findings are in full consistency with the results of e.g. \cite{traulsendet} (see e.g. Eqs. (7) and (8) of this reference), taking into account for example that $T^-(n)=0$ for all $n<n^*$, and that absorption occurs at the internal fixed point with probability one. Comparison with simulations confirms the validity of our calculations, see the right panel of Fig. \ref{fig:absorb1d}. As a technical detail we  point out that the time scale used in the figure is obtained from that of Eq. (\ref{eq:ttf}) via the re-scaling $t=\tau/N$. This is to ensure a well defined thermodynamic limit, $N\to\infty$, for example one may notice that the sum in (\ref{eq:ttf}) contains an ${\cal O}(N)$ number of terms, so that normalisation as just described is appropriate.


\section{Two-player symmetric games with cyclic interaction}\label{sec:imitrps}
\subsection{Internal absorption}

We will now turn to a different class of games, and consider symmetric two-players games with three pure strategies and cyclic interaction. Specifically we will investigate a generalisation of the well-known rock-paper-scissors (RPS) game, defined by the $3\times 3$ payoff matrix
\be\label{eq:rps}
A=\left(\begin{array}{ccc}
0 & 1 & -s \\
-s & 0 & 1 \\
1 & -s & 0 \end{array}
\right),
\ee
where we will generally consider the case $s>0$. Two players of identical species (strategies) will score zero payoff when they play each other. Strategy $1$ will `beat' strategy two (receiving a payoff of unity), but lose against strategy three, and similar for all cyclic permutations. The loser's payoff is $-s<0$. The central point of the $2$-dimensional strategy simplex, at which all three species are equally populated ($x_i=1/3$ for $i=1,2,3$) is known to be a stable fixed point of the corresponding replicator equation, provided $s<1$ \cite{gintis, claussen1, mobilia}. At $s=1$, corresponding to the standard RPS game, the centre is a neutrally stable fixed point, and the replicator dynamics admits an integral of motion, with periodic orbits about the centre fixed point. At $s>1$ finally the central fixed point is unstable, and deterministic trajectories approach the boundary of the simplex in a periodic manner. In absence of mutation, the case we will consider here, the three corner points of the simplex ($\bx=(1,0,0),(0,1,0),(0,0,1)$) are fixed points for any value of $s$.

The strict imitation dynamics in finite population is then defined by the following transition rates
\be\label{eq:transrps}
T_{i\to j}=\frac{n_i}{N}\frac{n_j}{N}\left(\frac{w}{2}\frac{\pi_j(\bn)-\pi_i(\bn)}{\Delta\pi_{\mbox{\tiny max}}}\right)\Theta\left(\pi_j(\bn)-\pi_i(\bn)\right), ~~i,j\in\{1,2,3\},~~i\neq j. 
\ee
We here use $\Delta\pi_{\mbox{\tiny max}}=1+s$ and $w=1$. 
\begin{figure}
\vspace{2em}
\centering
\includegraphics[scale=0.4]{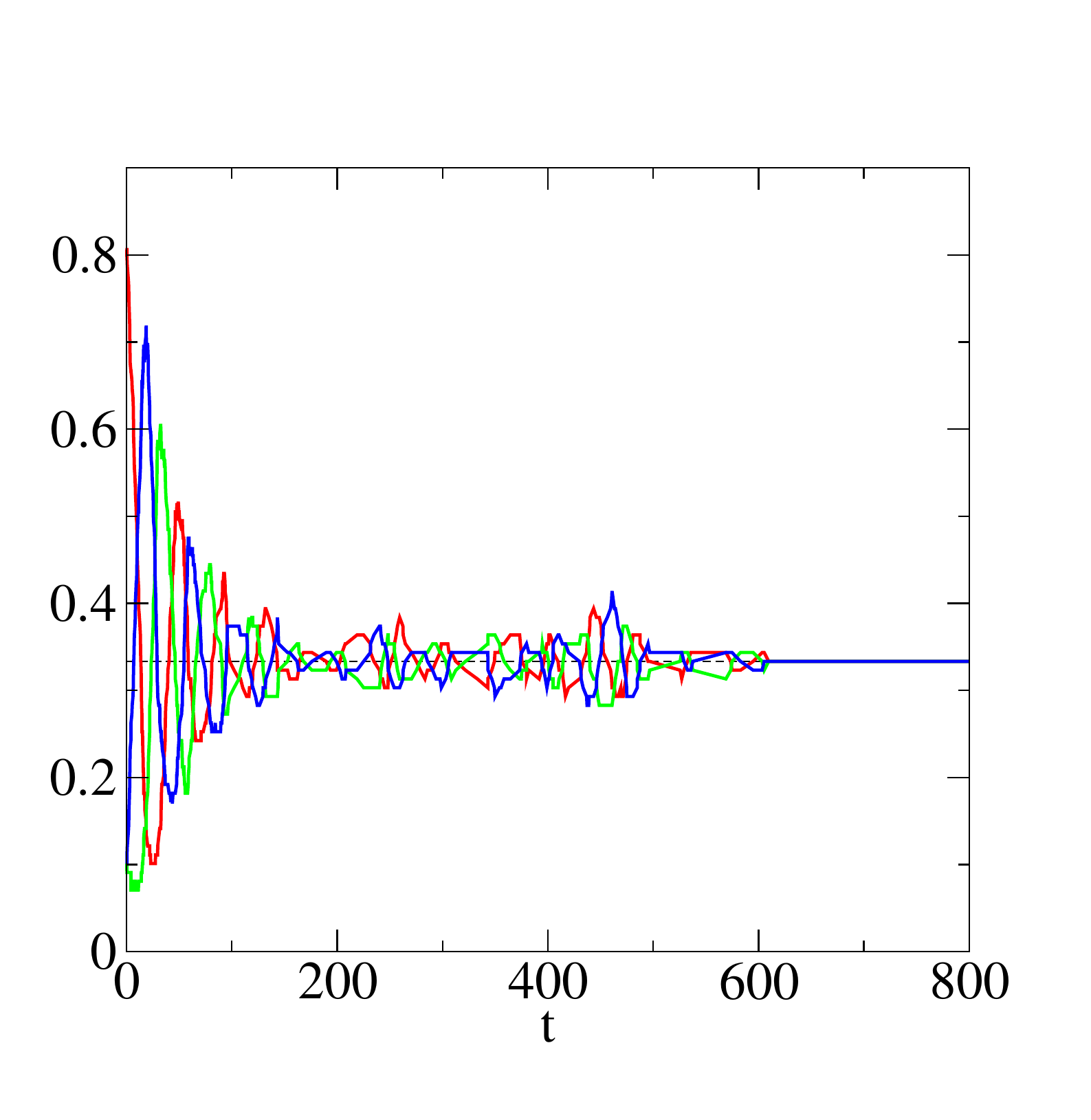}~~~~~\includegraphics[scale=0.4]{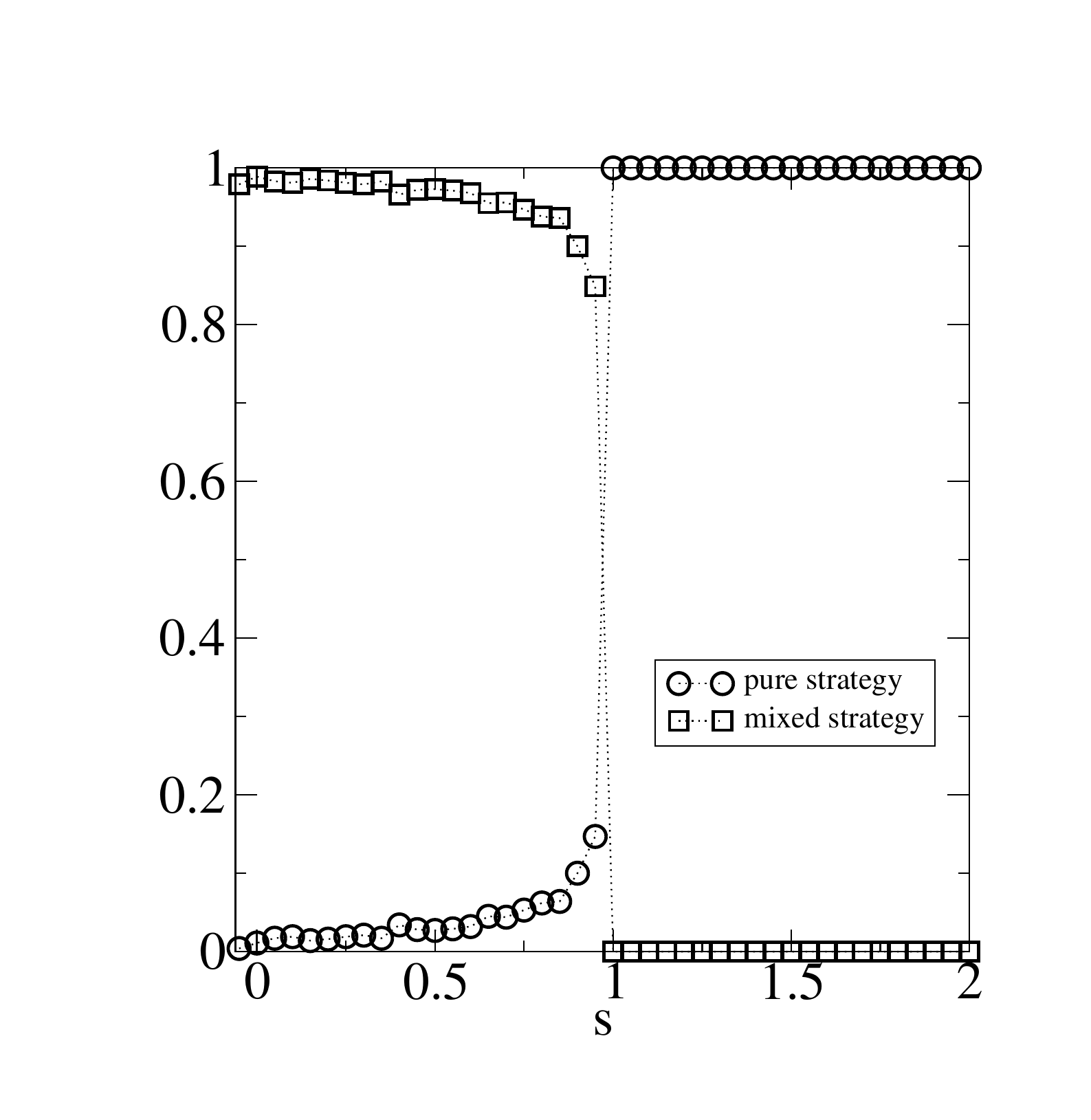}
\caption{(Color online) Left: Trajectory obtained for the generalised RPS game under strict proportional imitation ($N=99$, $s=0.5$). The the three curves show the concentrations of the three strategies as a function of time, convergence to the internal attractors is reached at about $t=600$. Right: Circles indicate the probability to fixate at one of the corners of the strategy simplex (pure strategies) if started from a random initial condition ($N=999$), squares show the probability of absorption at the central attractor (mixed strategy). Results are from simulations ($1000$ samples, run until fixation), and shown as a function of the model parameter $s$. }.
\label{fig:absorbrps}
\end{figure}

As shown in the left panel of Fig. \ref{fig:absorbrps}, this dynamics admits absorption at the internal fixed point. At the centre of the strategy simplex all three strategies are equally abundant, and by symmetry their fitnesses are equal. Hence all transition rates in Eq. (\ref{eq:transrps}) vanish. This is the case for any value of $s$, however as seen in the right panel of Fig. \ref{fig:absorbrps} fixation at the centre occurs with non-zero probability only at $s<1$. To understand this in more detail, it is helpful to keep in mind that the system can either fixate at the centre (corresponding to a mixed strategy profile) or at one of the corners of the strategy simplex (pure strategies). Due to stochastic fluctuations fixation at one of the four attractors will occur eventually in finite systems. This is confirmed in simulations, detailed analysis of the data in the right panel of Fig. \ref{fig:absorbrps} shows that all trajectories fixate eventually. Generally the stochastic dynamics will be governed by two contributions, first the flow of the deterministic limit about which the finite system can be thought to fluctuate, and secondly stochastic drift due to finite-size fluctuations. Stochasticity tends to drive the system to the boundaries of the simplex, and can hence cause fixation in one of the pure strategies (once the system hits one of the edges of the strategy simplex, fixation into one of the corners occurs). The results of Fig. \ref{fig:absorbrps} can then be understood as follows:
\begin{figure}
\vspace{2em}
\centering
\includegraphics[scale=0.6]{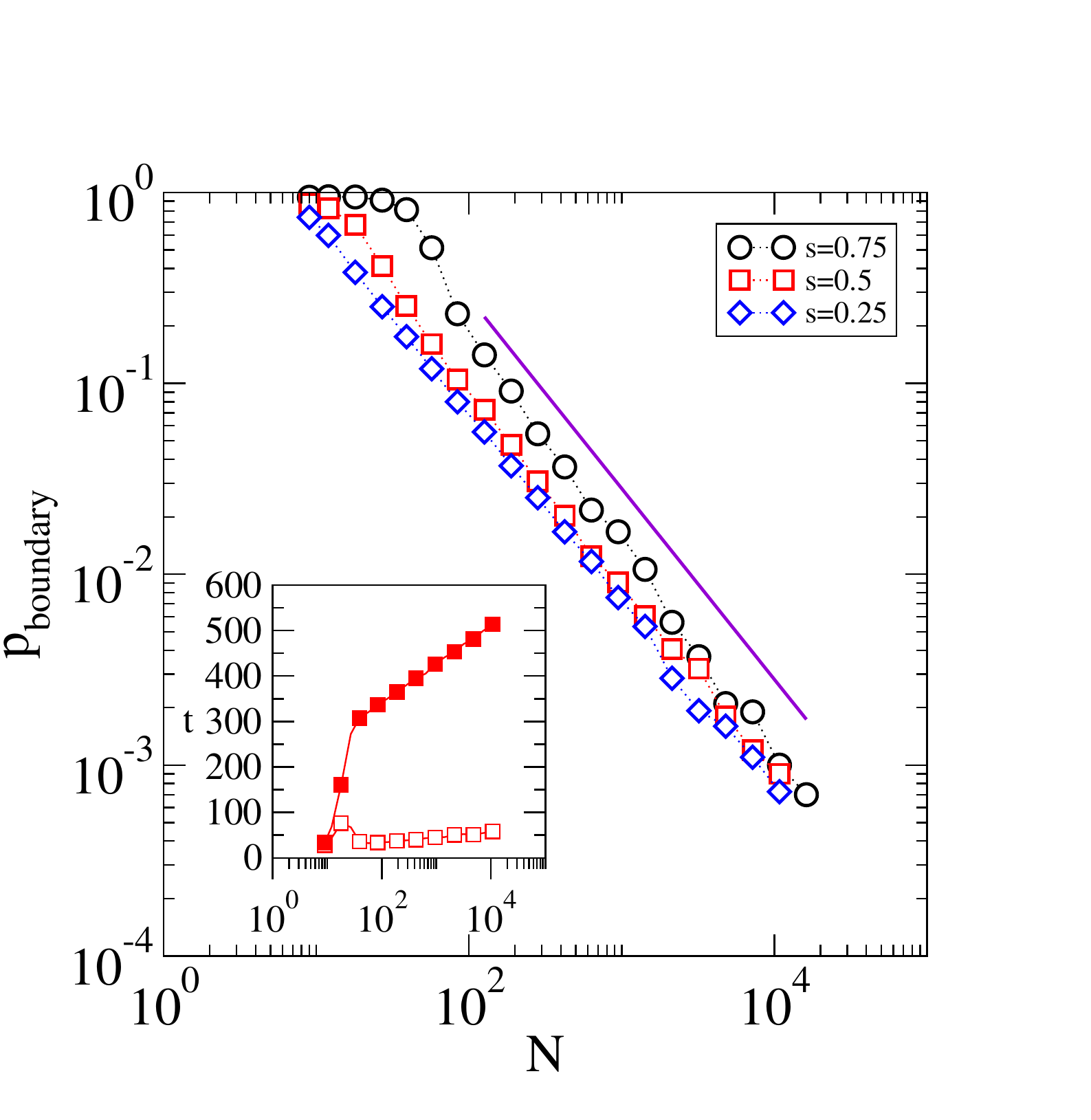} 
\caption{(Color online) Probability that fixation occurs at one of the corners of the strategy simplex in the generalise rock-paper-scissors game under strict proportional imitation dynamics ($\nu=0$). Results are shown as a function of the size $N$ of the population.  Symbols represent simulations, averaged over $10000-40000$ samples (depending on $s$), started from random initial conditions. The solid lines has a scaling of $1/N$. Inset: Average time to fixation at the boundary (open symbols) and the centre (filled symbols) for $s=0.5$}.
\label{fig:pboundary}
\end{figure}
\begin{enumerate}
\item[1.] At $s>1$ both the deterministic flow and the stochastic fluctuations drive the system towards the exterior of the simplex, and fixation occurs at one of the pure strategies.
\item[2.] At $s<1$ the deterministic flow pulls the system towards the centre of the strategy simplex, where fixation occurs, if the centre is reached. Stochastic fluctuations however provide a tendency to drift outwards, and provided one of the edges of the simplex is reached fixation occurs at a pure strategy. The deterministic pull towards the centre will be strong for values of $s$ much smaller than one, and fixation at the centre is likely. For values of $s$ close to one, the deterministic fixed point becomes less and less stable, and fluctuations more influential. The system fixates at the borders of the simplex with increasing probability, as seen in Fig. \ref{fig:absorbrps} (right panel), notice in particular the non-zero probability of fixation at a pure strategy for values of $s$ just below one.
\end{enumerate}
Measurements of the probability with which the system fixates at the boundary for fixed values of $s<1$, and at varying system sizes confirm that external absorption (i.e. fixation at one of the pure strategies) is indeed a stochastic effect induced by fluctuations. Fig. \ref{fig:pboundary} shows that the rate with which the system fixates at the edges of the simplex scales as $N^{-1}$ at large values of $N$. 

\begin{figure}
\vspace{2em}
\centering
\includegraphics[scale=0.45]{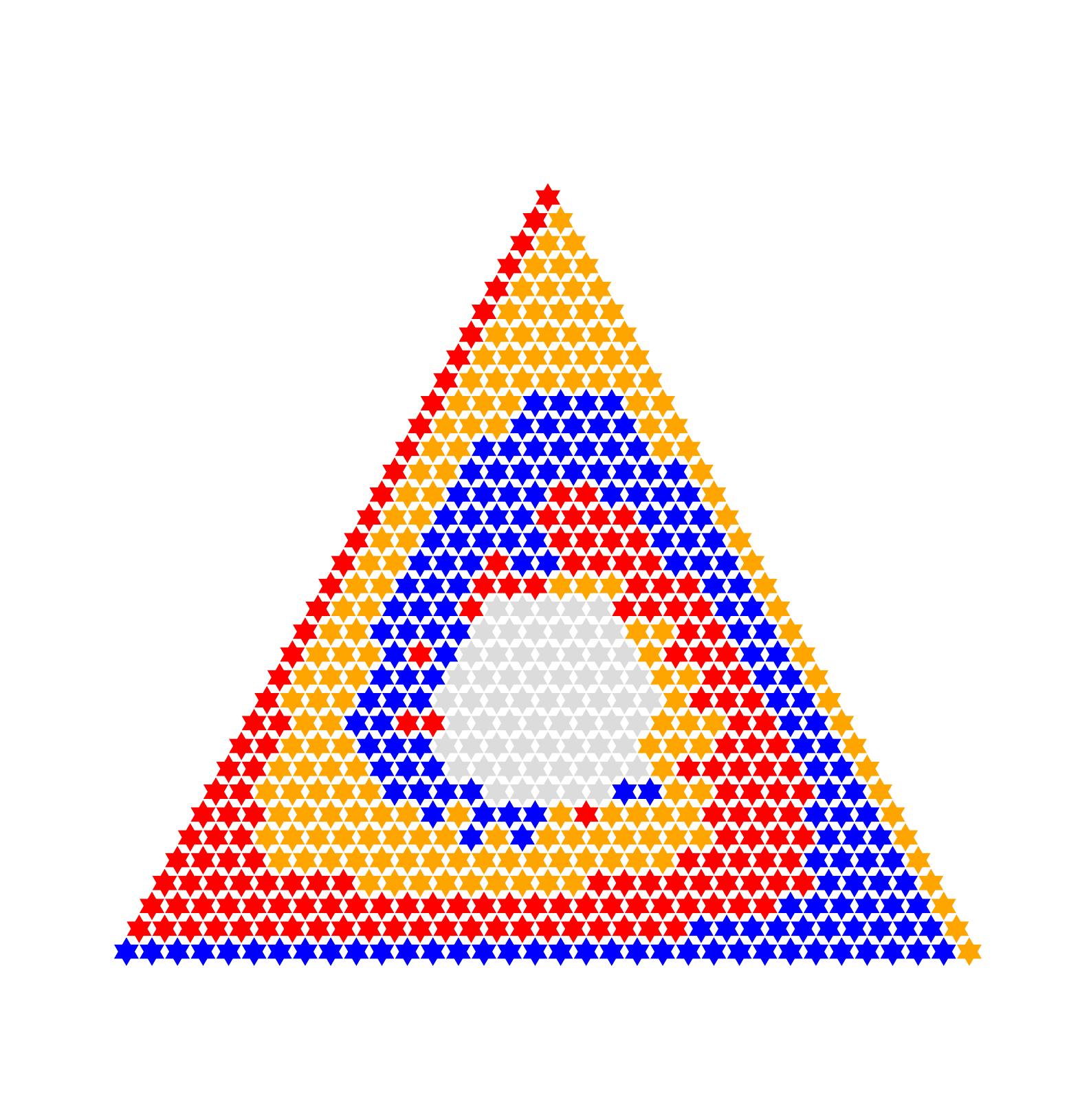}~~~~~~~~~\includegraphics[scale=0.45]{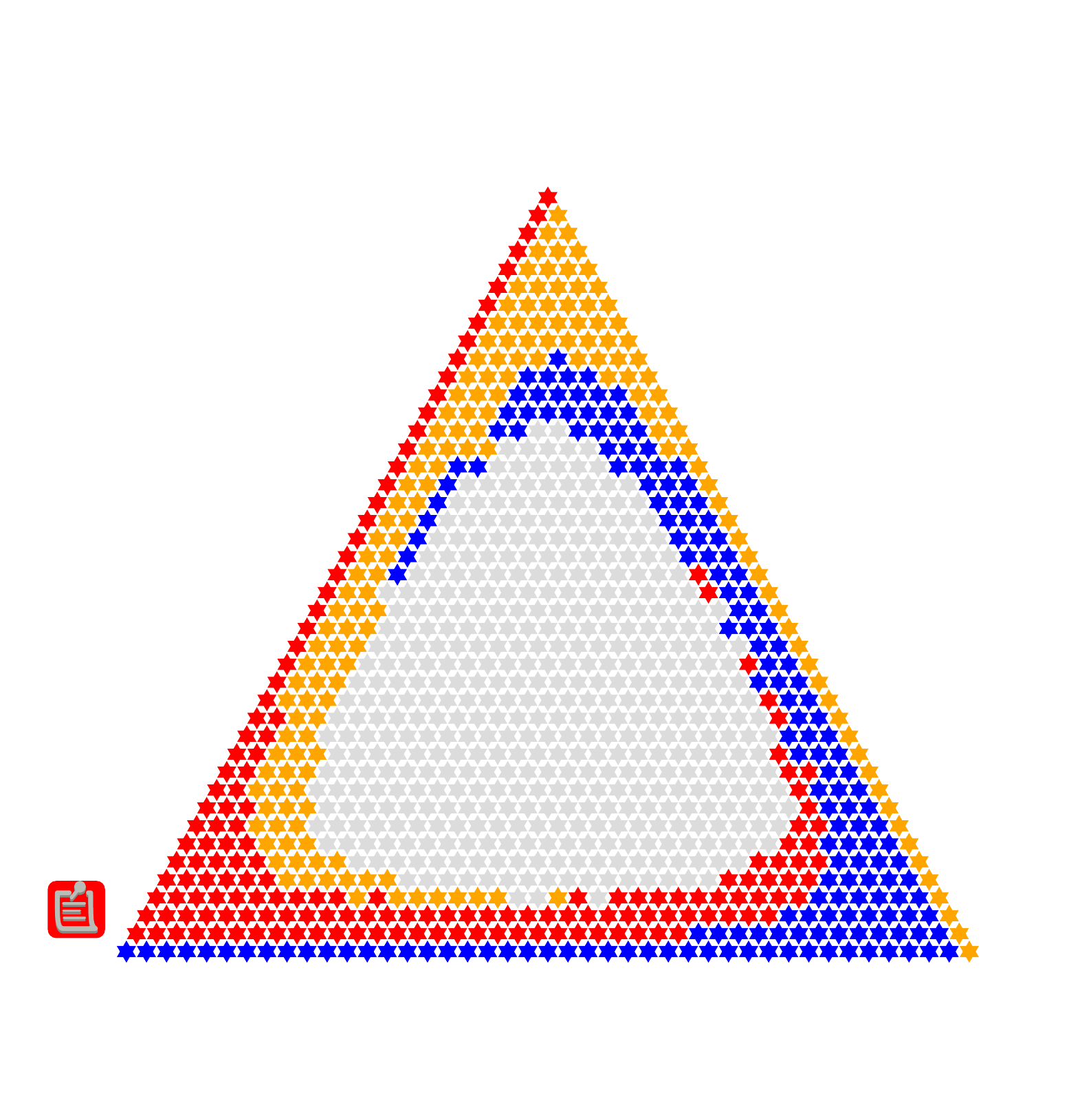}
\caption{(Color online) Fixation in the generalised rock-paper-scissors game at $s=0.75$ under strict adaption dynamics. The figure shows results from simulations indicating where the system is most likely to fixate given a starting position. Red symbols for example indicate fixation in the upper corner of the simplex. If started from the grey area near the centre the system is found to most likely fixate at the internal fixed point $(1/3,1/3,1/3)$. System sizes are $N=33$ (left) and $N=42$ (right). Simulations are run for in excess of $10^4$ independent samples per data point.}
\label{fig:rps_simplex}
\end{figure}

This can be further understood from Fig. \ref{fig:rps_simplex}, where we show what the most likely point of fixation is for different starting points of the dynamics. If started in the central region of the strategy simplex (indicated by grey symbols) it is more likely that the system will fixate at the centre than at any of the three pure strategy absorbing states. For starting points closer to the edges of the simplex, fixation is likely to occur at one of the corners, as indicated by the coloured symbols. The left and right panels compare the basins of attraction of the internal absorbing state for different system sizes. As seen in the figure the domain from which the central attractor is reached grows as the system size is increased, and essentially spans the bulk of the two-dimensional strategy simplex for sufficiently large systems. Fixation at the edges only occurs for starting points at the periphery. This leads to the conclusion that the results of Fig. \ref{fig:pboundary} may potentially be interpreted as an effect of bulk versus periphery. A starting point in the central region (which has a size ${\cal O}(N^2)$) favours fixation at the centre, a start in the peripheral region (which has a size ${\cal O}(N)$) predominantly leads fixation at one of the corners. At large $N$ and started from a random initial condition fixation occurs with a probability scaling with the relative areas of the peripheral and central basins of attraction, leading to a probability of external fixation scaling as $1/N$.  As a final remark we report that the average time to fixation at the interior fixed points (averaged over all initial conditions) is logarithmic in time (see inset of Fig. \ref{fig:pboundary}).

\subsection{Drift reversal}
We will now briefly turn to a discussion of local drift reversal in the generalised rock-papers-scissors game. As pointed out above such phenomena can only be expected when the considered dynamics has a stable attracting interior fixed point in the deterministic limit. Also we need to exclude internal absorption. We therefore focus on the above local process, and the generalised RPS game, Eq. (\ref{eq:rps}), with $s<1$.
\begin{figure}[t]
\vspace{2em}
\centering
\includegraphics[scale=0.5]{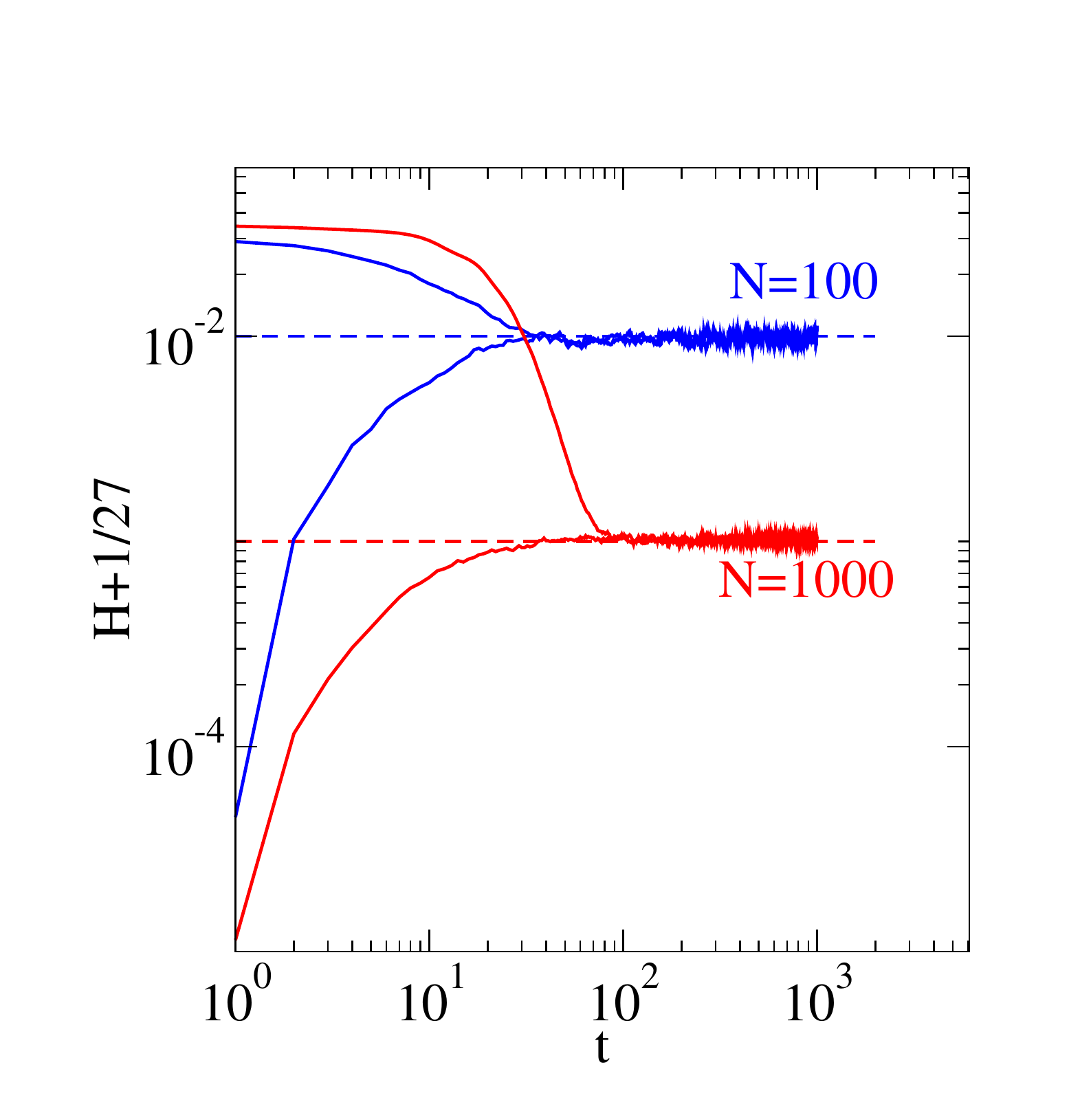}  
\caption{(Color online) Drift reversal in generalised RPS: The figure shows time series $H(t)-H^*$ from simulations of the local process at two different system sizes ($s=0.5$). In both cases the dynamics were once started from a fixed initial condition close to the central deterministic fixed point (increasing curves), and once from a point closer to one of the edges of the simplex (decreasing curves). Trajectories which fixate during the course of the simulation up to $t=10^3$ are excluded from the statistics, simulation data is from in excess of $300$ non-fixated runs. As seen in the figure the initial drift is away from the fixed point in the former case, and towards the fixed point in the latter. The dashed lines show the asymptotic value of $H-H^*$ as predicted by Eq. (\ref{eq:stath}). }
\label{fig:rpsdrift}
\end{figure}

The limiting replicator equations for the degrees of freedom $x_1,x_2$ and $x_3=1-x_1-x_2$, denoting the concentrations of the three strategies, are then known to admit the Liapunov function \cite{gintis, traulsen2}
\be
H=-x_1x_2(1-x_1-x_2).
\ee
Specifically one has $\dot H<0$ under the replicator dynamics ($s<1$), and the deterministic dynamics converges to the central fixed point $(x_1^*,x_2^*,x_3^*)=(1/3,1/3,1/3)$, corresponding to $H^*=-1/27$. In \cite{claussen1} the occurrence of drift reversal was reported for stochastic dynamics for this game in finite populations. Specifically, random initial conditions in the strategy simplex were considered, and it was shown that different microscopic processes tend to lead to an increase of $H$ in small populations (drift away from the central fixed point), and to decreasing values of $H$ in large populations (drift towards the fixed point). The `critical' population size, at which the drift reverses sign, can be computed within an expansion in the inverse system size, see \cite{claussen1,claussen2}. Crucially however, drift reversal was considered only as a global phenomenon, averaged over all possible initial conditions in strategy space. The main objective of this section is to refine this analysis, and to stress that drift and its reversal are again local phenomena. 

This can be observed for example in Fig. \ref{fig:rpsdrift}, where we show that the initial drift is towards the fixed point (decreasing curves in the figure), when the dynamics is started sufficiently far away from the centre of the simplex, but that a drift away from the centre (increasing time series of $H$) occurs for starting points close to the center.
\begin{figure}[t]
\vspace{2em}
\centering
 \includegraphics[scale=0.5]{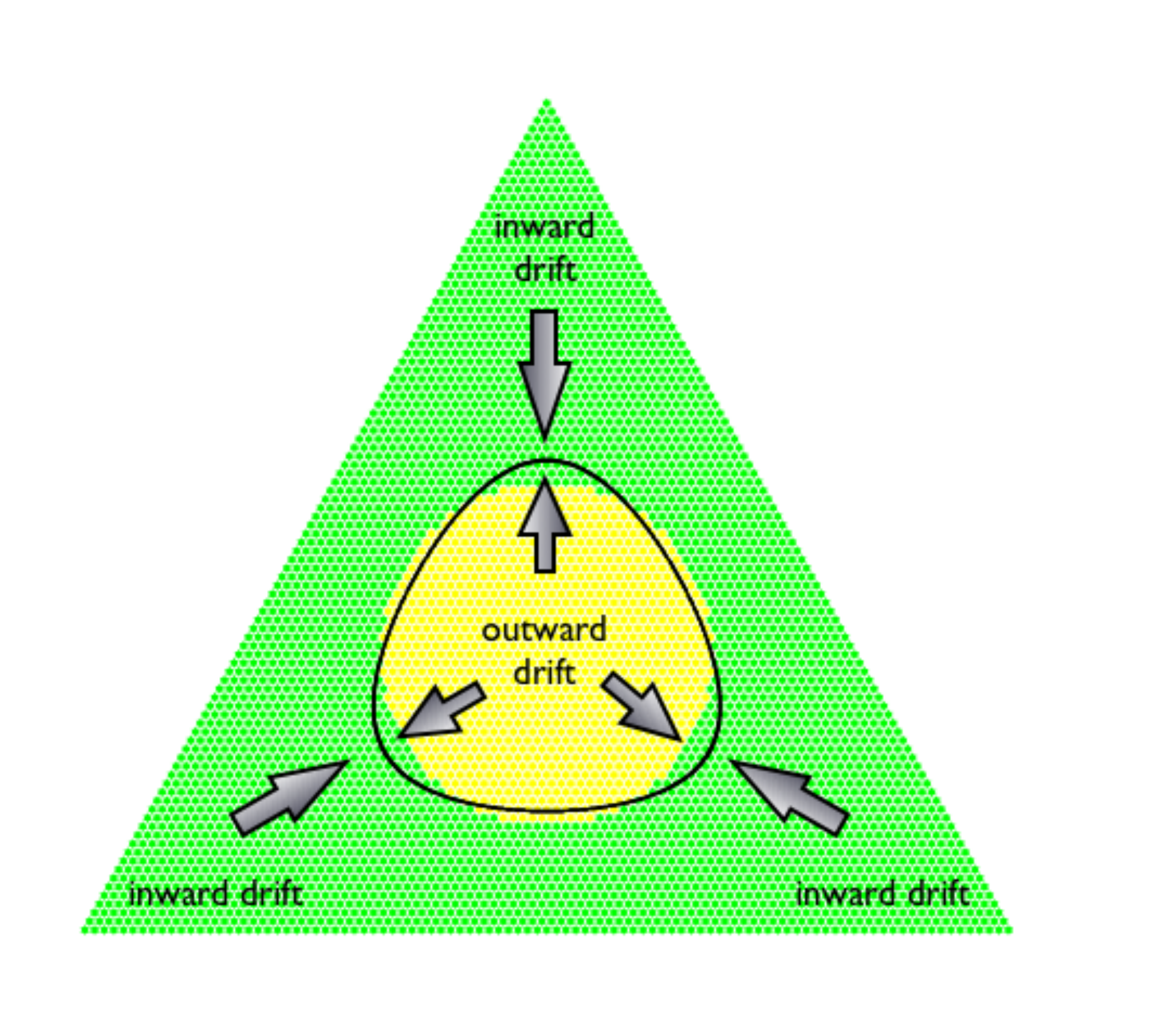}
\caption{(Color online) Drift reversal in a generalised RPS population ($s=0.5$) with $N=99$ individuals, subject to the local update process. Yellow symbols near the centre indicate the region in strategy space at which the drift has an outward direction, darker symbols in the outer region indicate an inward drift. These were obtained from numerically evaluating the sign of the drift, using an analogue of Eq. (\ref{eq:delta}). The solid black line indicates points with the value $H=\overline H$ predicted by Eq. (\ref{eq:stath}). Small deviations of the yellow region from the inside of the perimeter defined by $H=\overline H$ are due to finite-size effects.}
\label{fig:rpsdrift2}
\end{figure}
The simulation results in Fig. \ref{fig:rpsdrift} suggest that for any given population size $N$ the system chooses an asymptotic value, $\overline{H}$, of $H$, and that drift always occurs towards this preferred value of $H$.  Generally $\overline H$ will depend on the system size. Thus if, at a given value of $N$, the initial condition is such that the system is further away from the fixed point than indicated by $\overline{H}$, an inward drift occurs. Similarly, for starting points too close to the centre, an outward drift, together with an increase of $H$ is found, until $\overline{H}$ is reached. One should here stress that states with $H=\overline{H}$ are of course not absorbing, but that statements such as `the system will reach a value of $H=\overline{H}$', instead only hold statistically speaking, and when averaged over realisations of the stochastic dynamics.

Analytical progress can here be made within a system-size expansion, which allows one to describe the fluctuations about the deterministic fixed point. Again we do not discuss the detailed mathematics, as these are available in the literature, but report only the final result, consisting of the the following Langevin equations for fluctuations in the $x_1$ and $x_2$ components of the dynamics (fluctuations in the $x_3$ component then follow trivially due to overall conservation of particle number). One has
\BE
\dot \xi_1&=& \frac{w}{\Delta\pi_{\mbox{\tiny max}}}\left(\frac{s}{3}\xi_1+\frac{1+s}{3}\xi_2 \right)+ \eta_1 \nonumber \\
\dot \xi_2&=& \frac{w}{\Delta\pi_{\mbox{\tiny max}}}\left(-\frac{1+s}{3}\xi_1-\frac{1}{3}\xi_2\right)+\eta_2,\label{eq:langrps}
\EE
where $\xi_i$ are the fluctuations of the $x$-component, more precisely one has $n_i/N=x_i^*+\xi_i/\sqrt{N}$ for $i=1,2$. The variables $\eta_1(t)$ and $\eta_2(t)$ denote zero-average Gaussian white noise with the following covariance properties
\BE
\avg{\eta_1(t)\eta_1(t')}&=&\avg{\eta_2(t)\eta_2(t')}= \frac{2}{9}\delta(t-t') \\
\avg{\eta_1(t)\eta_2(t')}&=& -\frac{1}{9}\delta(t-t'), 
\EE
as obtained from the van Kampen expansion.
The linearity of Eqs. (\ref{eq:langrps}) allows further progress and in particular the long-time limit of the following quantities can be computed
\be
\sigma_{11}(t)=\avg{\xi_1(t)\xi_1(t)}, ~~ \sigma_{22}(t)=\avg{\xi_2(t)\xi_2(t)}, ~~\sigma_{12}(t)=\sigma_{21}(t)=\avg{\xi_1(t)\xi_2(t)},
\ee
see \cite{risken} for further details. Denoting these asymptotic values by $\overline\sigma_{11}, \overline\sigma_{22}$ and $\overline\sigma_{12}$ respectively, one then has
\BE
\overline H=-\avg{\left(x_1^*+\frac{\overline\xi_1}{\sqrt{N}}\right)\left(x_2^*+\frac{\overline\xi_1}{\sqrt{N}}\right)\left(1-x_1^*-x_2^*-\frac{\overline\xi_1}{\sqrt{N}} -\frac{\xi_2}{\sqrt{N}}  \right)},
\EE
where $\overline \xi_1$ and $\overline \xi_2$ are zero-average Gaussian random variables with statistics $\avg{\overline \xi_i\overline \xi_j}=\overline\sigma_{ij}$ for $i,j=1,2$, and where $\avg{\cdots}$ denotes an average over $\overline \xi_1$ and $\overline \xi_2$. From this one finds
\be\label{eq:stath}
\overline H-H^*=\frac{1}{3N}\left[\overline\sigma_{11}+\overline\sigma_{22}+\overline\sigma_{12}\right]
\ee
This estimate of the stationary value of $\overline H-H^*$, accurate for large, but finite $N$, compares well with simulations, see Fig. \ref{fig:rpsdrift}, and confirms the picture discussed above. Depending on the system size the dynamics tends to a stationary value $\overline H-H^*={\cal O}(N^{-1})$, see Eq. (\ref{eq:stath}). Interpreting $\overline H-H^*$ as a `distance' from the central deterministic fixed point, the dynamics is thus driven towards a state of a preferred distance from the central point. If started from closer to the fixed point than this distance then the initial drift is outwards, if started from a point outside the asymptotic perimeter set by $H=\overline H$, then the drift is inwards. This is illustrated in Fig. \ref{fig:rpsdrift2}. 

As a final technical and more subtle point we would like to add that the corners of the strategy simplex still remain absorbing states, as we are working in the absence of mutation. One should hence expect that any trajectory of the system in finite populations will fixate on a pure strategy eventually. While this may at first seem to be in contradiction with the results just discussed, it is important to keep in mind that this type of fixation occurs on an exponentially slow time scale for $s<1$, see e.g. \cite{mobilia}. This fixation regime is not captured by the van Kampen expansion. Instead the expansion addresses a regime in which a stationary Gaussian distribution is assumed, and where fixation at the edges of the simplex can be discarded. For the comparison of the theoretical predictions obtained from the van Kampen expansion against numerical data, we have therefore excluded trajectories which fixate in our simulations (see Fig. \ref{fig:rpsdrift}).

\section{Imitation dynamics in two-player $2\times 2$ games}\label{sec:imitbs}
In this section we will now consider a final example, and move on to a more complex asymmetric two-population game, Darwin's so-called `Battle of the Sexes', also known as the `Matching Pennies' game. Here there are two separate populations, each of size $N$, and labelled $1$ and $2$ for simplicity. Players in population $1$ (`male') only play the game with players in population $2$ (`female'), and vice versa. The payoff of a pure strategy in the male population is hence determined by the composition of the female population and vice versa. Payoffs can then be encoded in bi-matrix form as follows:
\be\label{eq:bspayoff}
\begin{array}{ccc}
& A_2 & B_2 \\
A_1 & (-1,+1) & (+1,-1) \\
B_1 & (+1,-1) & (-1,+1). 
\end{array}
\ee
The pure strategies in population $1$ are labelled $A_1$ and $B_1$, the pure strategies in population $2$ are referred to as $A_2$ and $B_2$. The first element in the brackets for each of the four cases is the payoff of the male player, the second the payoff of the female. In simple terms a male $A$-player prefers to interact with a female $B$-player rather than a female $A$, and a male $B$ preferably interacts with a female $A$ (anti-coordination). Female players take the opposite view, a female $A$ receives a higher payoff when meeting a male $A$ as compared to meeting a male $B$, and similarly a female $B$ prefers a male $B$ over a male $A$ (coordination). Of course the labelling `male' and `female' is not meant to have any direct real-world interpretation,  the game would be identical with these labels interchanged. We denote the number of male $A$ players by $n\in\{0,\dots,N\}$ and the number of female $A$ players by $m\in\{0,\dots,N\}$. The payoff structure is then described by fitness functions defined as follows
\BE
\pi^{(1)}_{A}(m)=-\frac{m}{N}+\frac{N-m}{N}, ~~~~ \pi^{(1)}_B(m)&=&-\pi^{(1)}_A(m), \nonumber \\
\pi^{(2)}_{A}(n)=\frac{n}{N}-\frac{N-n}{N},~~~~\pi^{(2)}_B(n)&=&-\pi^{(2)}_A(n).
\EE
Strict imitation dynamics occurs within the two populations, for example two male players may be chosen for potential adaptation, compare their fitnesses and then one of them may change their strategy to adopt that of the other male. This leads to the following transition rates
\BE
T^{(1),+}(n,m)&=&\frac{n}{N}\frac{N-n}{N}\left(\frac{w}{2}\frac{\pi^{(1)}_A(m)-\pi^{(1)}_B(m)}{\Delta\pi_{\mbox{\tiny max}}}\right)\Theta\left(\pi^{(1)}_A(m)-\pi^{(1)}_B(m)\right)\nonumber \\
T^{(1),-}(n,m)&=&\frac{n}{N}\frac{N-n}{N}\left(\frac{w}{2}\frac{\pi^{(1)}_B(m)-\pi^{(1)}_A(m)}{\Delta\pi_{\mbox{\tiny max}}}\right)\Theta\left(\pi^{(1)}_B(m)-\pi^{(1)}_A(m)\right)\nonumber \\
T^{(2),+}(n,m)&=&\frac{m}{N}\frac{N-m}{N}\left(\frac{w}{2}\frac{\pi^{(2)}_A(n)-\pi^{(2)}_B(n)}{\Delta\pi_{\mbox{\tiny max}}}\right)\Theta\left(\pi^{(2)}_A(n)-\pi^{(2)}_B(n)\right)\nonumber \\
T^{(2),-}(n,m)&=&\frac{m}{N}\frac{N-m}{N}\left(\frac{w}{2}\frac{\pi^{(2)}_B(n)-\pi^{(2)}_A(n)}{\Delta\pi_{\mbox{\tiny max}}}\right)\Theta\left(\pi^{(2)}_B(n)-\pi^{(2)}_A(n)\right).\label{eq:bsimit}
\EE
\begin{figure}
\vspace{2em}
\centering
\includegraphics[scale=0.5]{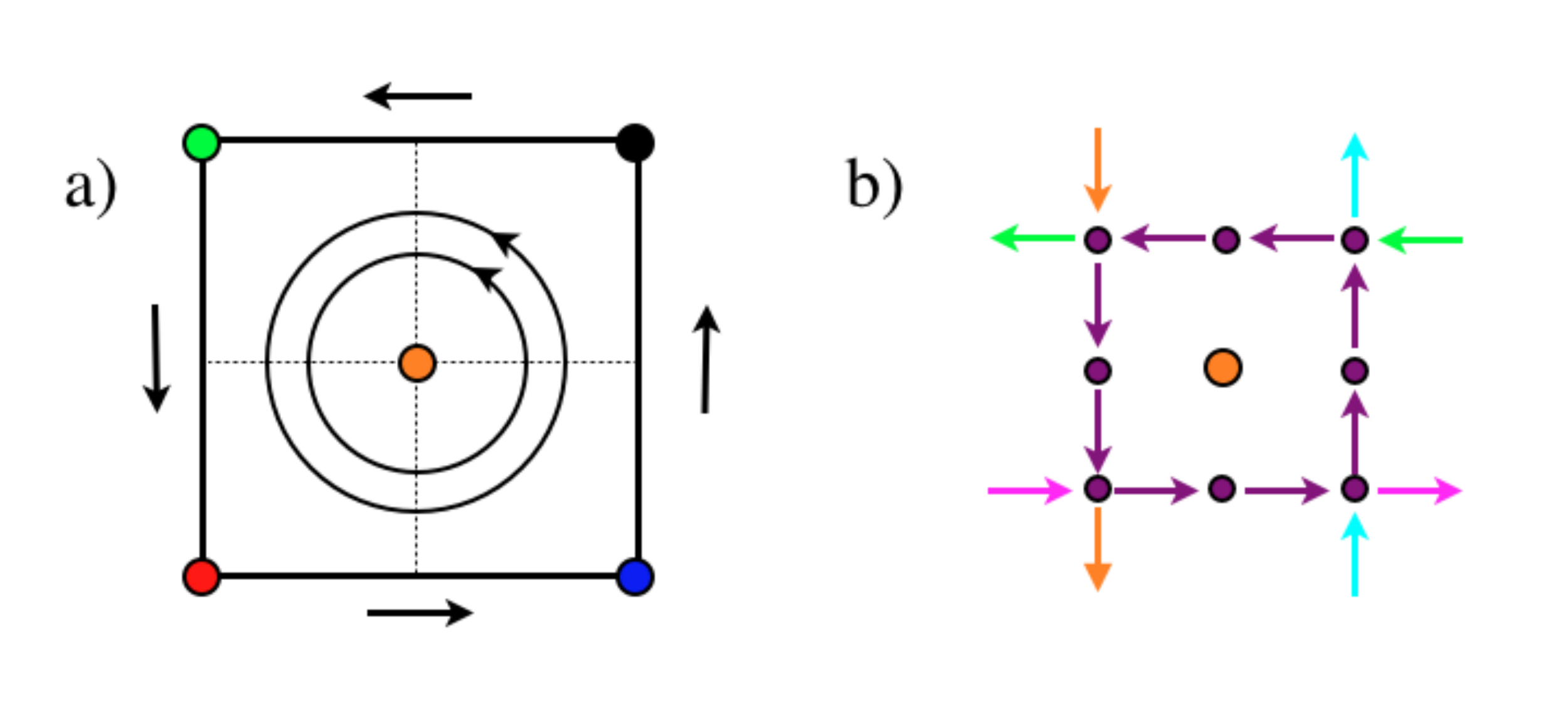}
\caption{(Color online) Battle of the Sexes: Panel a) shows an illustration of the deterministic flow, and the five fixed points. Panel b) shows the region near the central fixed point, arrows indicate the transitions that are possible under the strict imitation dynamics. The central fixed point can never be reached. The arrows into and out of the central square have been colored in pairs. The two arrows of any given color will give a net outflow taking into account that transition rates scale in proportion to ratios of the type $m(N-m)/N^2$ for example. }
\label{fig:bsflow}
\end{figure}

\begin{figure}
\vspace{2em}
\centering
\includegraphics[scale=0.5]{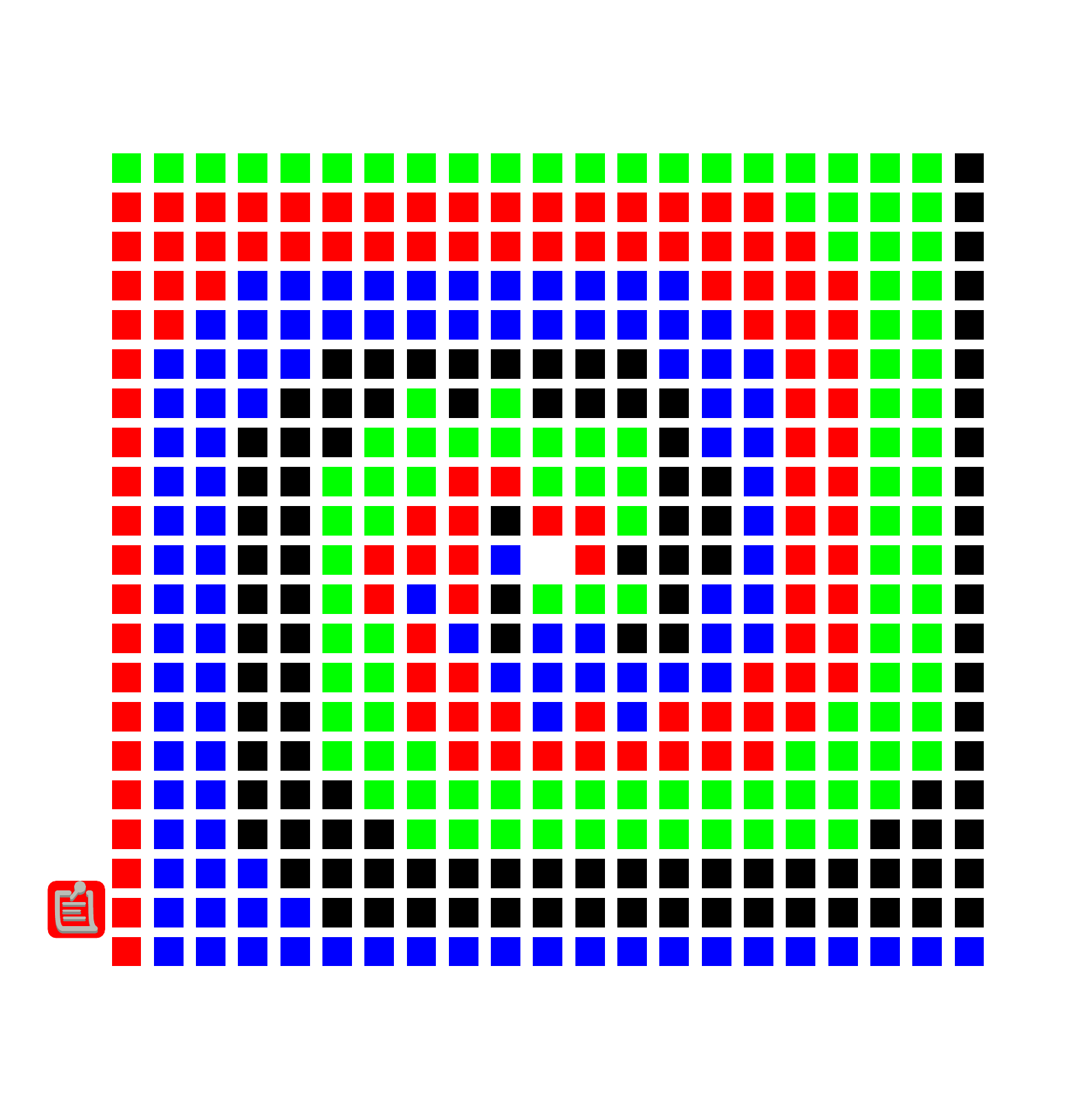}
\caption{(Color online) Fixation in the battle of the sexes under strict imitation dynamics. Colors indicate the absorbing state that is most likely to be reached if the dynamics is started from a given initial condition. Simulations are for $N=20$, $100001$ realisations. The central fixed point at $(n,m)=(10,10)$ is left blank.}
\label{fig:absorbbs}
\end{figure}

The resulting flow in the deterministic replicator limit is shown in Fig. \ref{fig:bsflow} (panel a), and has the well known cyclic motion with neutrally stable cycles about the central fixed point at $(1/2,1/2)$. While this point is an absorbing state under the strict imitation dynamics in finite populations, Eqs. (\ref{eq:bsimit}), simulations reveal that fixation never occurs at the centre, but that instead all trajectories are absorbed at one of the four corners of strategy space eventually. This is illustrated in Fig. \ref{fig:absorbbs}, where we indicate the most likely point of absorption for all possible initial conditions  of the imitation dynamics for $N=21$. The absence of internal absorption in this game can be understood from a closer inspection of the imitation dynamics close to the central fixed point (see Fig. \ref{fig:bsflow} panel b). In this figure we show the central fixed point (centre circle), and the surrounding states in finite populations (the population size is assumed to be even, so that the central fixed point is within the allowed strategy space). The arrows indicate transitions from one state to another that are possible under imitation, if no arrow is present the corresponding transition is not admissible. As seen in the figure the central fixed point cannot be reached from any of the neighbouring states, hence internal absorption can not occur. A closer analysis shows that there is indeed a net flux away from the region of the central fixed point, see the caption of Fig. \ref{fig:bsflow}.

\section{Conclusions}
In summary we have studied a strict imitation dynamics in games played by finite populations of players. It is here important to stress that strict imitation is at best a stylised dynamical process, and in the light of recent experiments \cite{traulsenpnas} may well be ruled out for real-word adaptation and learning of players. Still this dynamics has interesting dynamical features, as underlined by the early studies of Helbing \cite{helbing}, linking imitation to replicator dynamics. We have here shown that proportional imitation can give rise to what we call internal absorption, a novel dynamical phenomenon in which dynamic arrest occurs at an internal point in strategy space, corresponding to the stable attractor of the deterministic replicator dynamics. This effect occurs in simple symmetric $2\times 2$ games, provided the payoff matrix is chosen such that the replicator equations have a stable internal fixed point. We have also verified the occurrence of internal absorption in generalise rock-paper-scissors games, again the stability of the internal fixed point under the replicator dynamics appears to be a necessary condition. In the Battle of the Sexes, where the replicator equations have neutrally stable cycles, no absorption at the internal fixed point is found.  

The second main contribution of this paper is to analyse in detail the phenomenon of drift reversal, first reported in \cite{claussen1,claussen2}. We have shown that the reversal of drift is in fact a local phenomenon, dependent on position in strategy space. Provided its deterministic limiting dynamics has a stable internal fixed point, a given microscopic update rule will select a `typical' distance from this fixed point at which it operates asymptotically. This distance will depend on the population size, and will be smaller for large populations than for small ones. If started  closer to the fixed point than the asymptotic distance, the system will have a tendency to move away from the fixed point (outward drift), if started further away from the fixed point than the preferred distance, an inward drift will be observed. The precise domains of inward and outward drift can be obtained analytically within a system-size expansion, valid for large, but finite populations. In this limit we have also provided an analytical estimate for the typical asymptotic distance from the deterministic fixed point.


\begin{acknowledgments} 
The author acknowledges funding by the Research Councils UK (RCUK reference EP/E500048/1), and would like to thank A. Traulsen and J. C. Claussen for useful discussions.
\end{acknowledgments}


\section*{Appendix}
The point in strategy space at which drift reversal occurs for the one-dimensional Hawk-Dove example can also be obtained from a Kramers-Moyal expansion of the master equation (\ref{eq:master}). As shown in \cite{traulsen2,traulsen} this formalism results in a multiplicative stochastic process of the form
\be
dx=F(x)dt+\frac{G(x)}{\sqrt{N}}dW,
\ee
where $dW$ denotes the increment of a standard Wiener process. One has $F(x)=T^{\infty,+}(x)-T^{\infty,-}(x)$ and $G(x)=\sqrt{T^{\infty,+}(x)+T^{\infty,-}(x)}$. From this one then finds (respecting the rules of Ito calculus)
\BE
dD&=&D'(x)dx+\frac{1}{2}D''(x)(dx)^2 
\EE
for the time evolution of the distance $D$ from the interior fixed point. This then gives
\BE
\avg{dD(x)}&=&(D^\infty)'(x)F(x)dt+\frac{1}{2N}(D^\infty)''(x)(G(x))^2dW^2 \nonumber \\
&=&[(D^\infty)'(x)F(x)+\frac{1}{2N}(D^\infty)''(x)(G(x))^2]dt,
\EE
i.e.
\be
\frac{d}{dt}\avg{D|x}=(D^\infty)'(x)F(x)+(D^\infty)''(x)\frac{1}{2}\frac{G(x)^2}{N}
\ee
The notation $d\avg{D|x}/dt$ here indicates the expected change in distance per unit time at time $t$ conditioned to trajectories starting at position $x$ at $t$. The stationary point, $d\avg{D|x_s}/dt=0$ is therefore given by the condition $(D^\infty)'(x_s)F(x_s)=-(D^\infty)''(x_s)\frac{G(x_s)^2}{2N}$.
For the Euclidean distance $D^\infty(x)=(x-x^*)^2$ this simplifies to $ 2(x_s-x_*)F(x_s)=-\frac{G(x_s)^2}{N}$, and finally assuming that $x_s - x^*$ is of order $N^{-1/2}$, we self-consistently find
\be
x_s-x_*=\pm\sqrt{-\frac{G(x^*)^2}{2NF'(x^*)}}.
\ee
For the case of the Hawk-Dove game with local dynamics we find $F'(x^*)=-1/4$ and $G(x^*)^2=1/4$ so that we have
\be
x_s = \frac{1}{2}\pm\sqrt{\frac{1}{2N}},
\ee
reproducing the result of Eq. (\ref{eq:driftest}).

\end{document}